# Title: Associations between depression symptom severity and daily-life gait characteristics derived from long-term acceleration signals in real-world settings

Yuezhou Zhang[1], MSc; Amos A Folarin[1,2,3,15,16], PhD; Shaoxiong Sun[1], PhD; Nicholas Cummins[1], PhD; Srinivasan Vairavan[13], PhD; Linglong Qian[1], MSc; Yatharth Ranjan[1], MSc; Zulqarnain Rashid[1], PhD; Pauline Conde[1], BSc; Callum Stewart[1], MSc; Petroula Laiou[1], PhD; Heet Sankesara[1], BSc; Faith Matcham[4], PhD; Katie M White[4], BSc; Carolin Oetzmann[4], MSc; Alina Ivan[4], MSc; Femke Lamers[5], PhD; Sara Siddi[6,7,8], PhD; Sara Simblett[9], PhD; Aki Rintala[10,11], PhD; David C Mohr[12], PhD; Inez Myin-Germeys[10], PhD; Til Wykes[9,17], PhD; Josep Maria Haro[6,7,8], PhD; Brenda WJH Penninx[5], PhD; Vaibhav A Narayan[13], PhD; Peter Annas[14], PhD; Matthew Hotopf[3,4], PhD; Richard JB Dobson[1,2,3,15,16], PhD; RADAR-CNS consortium[18]

[1]Department of Biostatistics & Health Informatics, Institute of Psychiatry, Psychology and Neuroscience, King's College London, London, United Kingdom
[2]Institute of Health Informatics, University College London, London, United Kingdom
[3] NIHR Maudsley Biomedical Research Centre, South London and Maudsley NHS Foundation Trust, London, United Kingdom
[4]Department of Psychological Medicine, Institute of Psychiatry, Psychology and Neuroscience, King's College London, London, United Kingdom
[5]Department of Psychiatry, Amsterdam Public Health Research Institute and Amsterdam Neuroscience, Amsterdam University Medical Centre, Vrije Universiteit and GGZ inGeest, Amsterdam, Netherlands
[6]Teaching Research and Innovation Unit, Parc Sanitari Sant Joan de Déu, Fundació Sant Joan de Déu, Barcelona, Spain
[7]Centro de Investigación Biomédica en Red de Salud Mental, Madrid, Spain
[8]Faculty of Medicine and Health Sciences, Universitat de Barcelona, Barcelona, Spain
[9]Department of Psychology, Institute of Psychiatry, Psychology and Neuroscience, King's College London, London, United Kingdom
[10]Center for Contextual Psychiatry, Department of Neurosciences, Katholieke Universiteit Leuven, Leuven, Belgium
[11]Faculty of Social Services and Health Care, LAB University of Applied Sciences, Lahti, Finland
[12]Center for Behavioral Intervention Technologies, Department of Preventive Medicine, Northwestern University, Evanston, Illinois, United States
[13]Janssen Research and Development LLC, Titusville, NJ, United States
[14] H. Lundbeck A/S, Copenhagen, Denmark
[15] Health Data Research UK London, University College London, London, United Kingdom
[16] NIHR Biomedical Research Centre at University College London Hospitals NHS Foundation Trust, London, United Kingdom

[17] South London and Maudsley NHS Foundation Trust, London, United Kingdom
[18] https://radar-cns.org/

**Corresponding author**
Richard Dobson
Department of Biostatistics & Health Informatics
SGDP Centre, IoPPN
King's College London
Box PO 80
De Crespigny Park, Denmark Hill
London
SE5 8AF
Email: richard.j.dobson@kcl.ac.uk
Telephone: +44(0) 20 7848 0473

# Abstract

**Background**: Gait is an essential manifestation of depression. Laboratory gait characteristics have been found to be closely associated with depression. However, the gait characteristics of daily walking in real-world scenarios and their relationships with depression are yet to be fully explored.

**Objective:** This study aimed to explore associations between depression symptom severity and daily-life gait characteristics derived from acceleration signals in real-world settings.

**Methods:** In this study, we used two ambulatory datasets: a public dataset with 71 elder adults' 3-day acceleration signals collected by a wearable device, and a subset of an EU longitudinal depression study with 215 participants and their phone-collected acceleration signals (average 463 hours per participant). We detected participants' gait cycles and force from acceleration signals and extracted 20 statistics-based daily-life gait features to describe the distribution and variance of gait cadence and force over a long-term period corresponding to the self-reported depression score.

**Results:** The gait cadence of faster steps (75th percentile) over a long-term period has a significant negative association with the depression symptom severity of this period in both datasets. Daily-life gait features could significantly improve the goodness of fit of evaluating depression severity relative to laboratory gait patterns and demographics, which was assessed by likelihood-ratio tests in both datasets.

**Conclusion:** This study indicated that the significant links between daily-life walking characteristics and depression symptom severity could be captured by both wearable devices and mobile phones. The gait cadence of faster steps in daily-life walking has the potential to be a biomarker for evaluating depression severity, which may contribute to clinical tools to remotely monitor mental health in real-world settings.

# Introduction

Depression affects over 300 million people's lives worldwide [1] and is associated with many adverse outcomes, including decreased quality of life, loss of occupational function, disability, premature mortality, and suicide [2-5]. While early treatment can be effective and prevent more serious adverse outcomes [6], more than half of depressed people do not receive timely treatment [7, 8]. Current questionnaire-based depression assessments may be affected by recall bias and may not be able to collect dynamic information [9, 10]. Therefore, several recent studies have attempted to explore the associations between depression and changes in individuals' behaviors using mobile technologies [11].

Changes in gait are essential manifestations of depression [12, 13]. The main hypothesis linking gait with depression is a bi-directional interaction between the brain motor system and cortical and subcortical structures, which are related to emotions and cognitive functions [14-16]. Many studies have explored the relationships between depression and gait characteristics based on "gold standard" laboratory walking tests. Longer gait cycles, reduced stride length, and slower gait cadence were observed in participants with depression compared with healthy controls, which have been consistently shown in several studies [17-25]. Other gait abnormalities, such as the reduced gait force [21], increased double support time [22], reduced swing time variability [23], slumped postures [24], and increased body sway [25], have been reported, but with less consistency across studies.

Laboratory tests are hard to apply in real-world settings because of the need for expensive equipment (e.g. video camera and force plates), specialized laboratories, and the inconvenience of wearing sensors on knees and ankles for example [14, 26]. Further, studies have found that laboratory and daily-life gait measurements do not correlate perfectly probably due to subjective psychological factors, laboratory-controlled conditions, and complex situations in daily-life walking [27, 28]. Some researchers have suggested that people's daily-life activity characteristics should have stronger links to their health conditions than laboratory tests [27]. Therefore, it is necessary to monitor and evaluate daily-life walking using an efficient method.

With the development of sensor technology, steps in daily-life walking can be detected by mobile phones and convenient wearable devices, which provide a cost-efficient, continuous, and unobtrusive means to quantitatively monitor daily-life gait characteristics. In recent years, several studies have used mobile phones or wearable devices for long-term monitoring of daily-life gait characteristics and explored their links with fall risks [28, 29] and neurological disorders [30]. Weiss et al found that accelerometer-derived measures based on long-term monitoring improved the identification of fall risks compared with laboratory tests [29]. A recent review stated that free-living monitoring using an accelerometer can confer advantages over clinical assessments in Parkinson's disease [30]. However, for the relationships between depression and daily-life walking, to the best of our knowledge, only the number of steps has been investigated [31-33]. The number of steps is more focused on reflecting levels of individuals' mobility and physical activity, rather than gait characteristics. Gait characteristics of daily-life walking, such as gait cadence, gait force, and variance in gait, and their associations with depression are yet to be fully explored.

To fill this gap, this study aimed to explore the associations between depression symptom severity and daily-life gait characteristics derived from mobile technologies. Specifically, we extracted several features related to gait cadence [34] and gait force [35] which could be extracted from acceleration signals to represent characteristics of daily-life walking over a long-term period, and assessed their associations with corresponding self-reported depression scores of this period. We also tested whether daily-life gait characteristics could provide additional value for evaluating depression symptom severity relative to laboratory gait patterns or demographics. To explore whether associations between daily-life walking and depression could be captured by different accelerometer devices, we performed our analyses on two ambulatory datasets whose acceleration signals were collected by a wearable device and mobile phone, respectively [29, 36].

# Methods

## Datasets

### Long Term Movement Monitoring dataset

The Long Term Movement Monitoring (LTMM) dataset includes 71 elderly adults' demographics (age and gender), depression scores (the 15-item Geriatric Depression Scale [GDS-15]), and acceleration signals of laboratory walking tests and 3-day activities [29], which can be downloaded at PhysioNet [37]. None of the participants in the LTMM dataset showed signs of cognitive impairment (according to the Mini-Mental State Examination) and any gait or balance disorders [29]. At the enrollment session, the participant's depression symptom severity was estimated using the GDS-15 whose total score ranges from 0 to 15 (increasing depression symptom severity) with a cutoff score of $\geq 5$ indicating probable depressive disorders [38]. Participants were asked to walk at a self-selected and comfortable speed for 1 minute in the laboratory while wearing a 3-axis accelerometer on their lower back [29]. After the laboratory walking test, all participants were asked to wear the accelerometer for the next 3 consecutive days to record daily activities. All acceleration signals were recorded at 100 Hz [29].

### RADAR-MDD-KCL dataset

The EU research program Remote Assessment of Disease and Relapse – Major Depressive Disorder (RADAR-MDD) aimed to investigate the utility of mobile technologies for long-term monitoring of participants with depression in real-world settings [36, 39]. In this paper, we used a subset of RADAR-MDD which was collected from a study site in the United Kingdom (King's College London [KCL]), because the KCL site was the only site to acquire ethical approval for collecting the phone's acceleration signals of participants' daily-life walking. We denoted this subset as the RADAR-MDD-KCL dataset for reading convenience. The phone's acceleration signals were collected at 50 Hz and uploaded to an open-source platform, RADAR-base [40]. The participant's depression symptom severity was assessed by the 8-item Patient

Health Questionnaire (PHQ-8) conducted through mobile phones every 2 weeks. The total score of the PHQ-8 ranges from 0 to 24 (increasing severity) with a cutoff score $\geq 10$ for clinically significant depression symptoms [41]. Participants' demographics (age and gender) and the number of comorbidities (Supplement Table 1) were considered as covariates in this study to control confounding factors that may affect gait characteristics. A patient advisory board comprising service users co-developed the study. They were involved in the choice of measures, the timing, and issues of engagement and have also been involved in developing the analysis plan.

**Step detection algorithm**

Since we need to respectively detect steps on the acceleration signals collected by wearable devices and mobile phones, we chose to apply the step detection algorithm presented in [42], which was based on mobile phones (Figure 1). Given a segment of 3-axis acceleration signals $(x_i, y_i, z_i)$, first, the magnitude of the acceleration of the segment of acceleration signals was calculated to combine 3-dimension signals to a single series $r_i$, where $r_i = \sqrt{x_i^2 + y_i^2 + z_i^2}$. The magnitude of the acceleration signals does not depend on the orientation and tilt of the mobile phone during walking [42]. Then, $r_i$ was filtered by a weighted moving average filter to remove noise (equation 1, $\omega = 150ms$). Next, the filtered $\bar{r}_i$ was subtracted by the mean of $\bar{r}_i$ to make $\bar{r}_i$ symmetric to the x-axis. We calculated two new series $B1_i$ and $B2_i$ based on two thresholds to detect the walking swing phase and stance phase, respectively (Equation 2 and 3). If a swing phase ends and a stance phase starts, we can identify a step that occurred. The formal detection rule of a step $S_i$ at sample $i$ is the following two conditions must be satisfied: 1) a change from -0.5 to 0 in B1 ($B1_i = 0$ and $B1_{i-1} = 0.5$); 2) there is at least one detection of $B2 = -0.5$ in a window of size $\omega = 150ms$ after sample $i$ ($Min(B2_{i:i+\omega}) = -0.5$).

$$\bar{r}_i = \frac{1}{2\omega+1}\sum_{j=i-\omega}^{i+\omega} r_j \quad (1)$$

$$B1_i = \begin{cases} 0.5, & if\ \bar{r}_i \geq 0.5 \\ 0, & otherwise \end{cases} \quad (2)$$

$$B2_i = \begin{cases} -0.5, & if\ \bar{r}_i \leq -0.5 \\ 0, & otherwise \end{cases} \quad (3)$$

**Gait cycles and gait force**

Then, the gait cycle series could be derived by calculating time intervals between consecutive steps, which was denoted as $Cycles$. During each gait cycle, the amplitude from peak to valley of the magnitude of the acceleration signals was used to reflect the gait force of each step. The force of all steps in the given acceleration signal was denoted as the series $Force$.

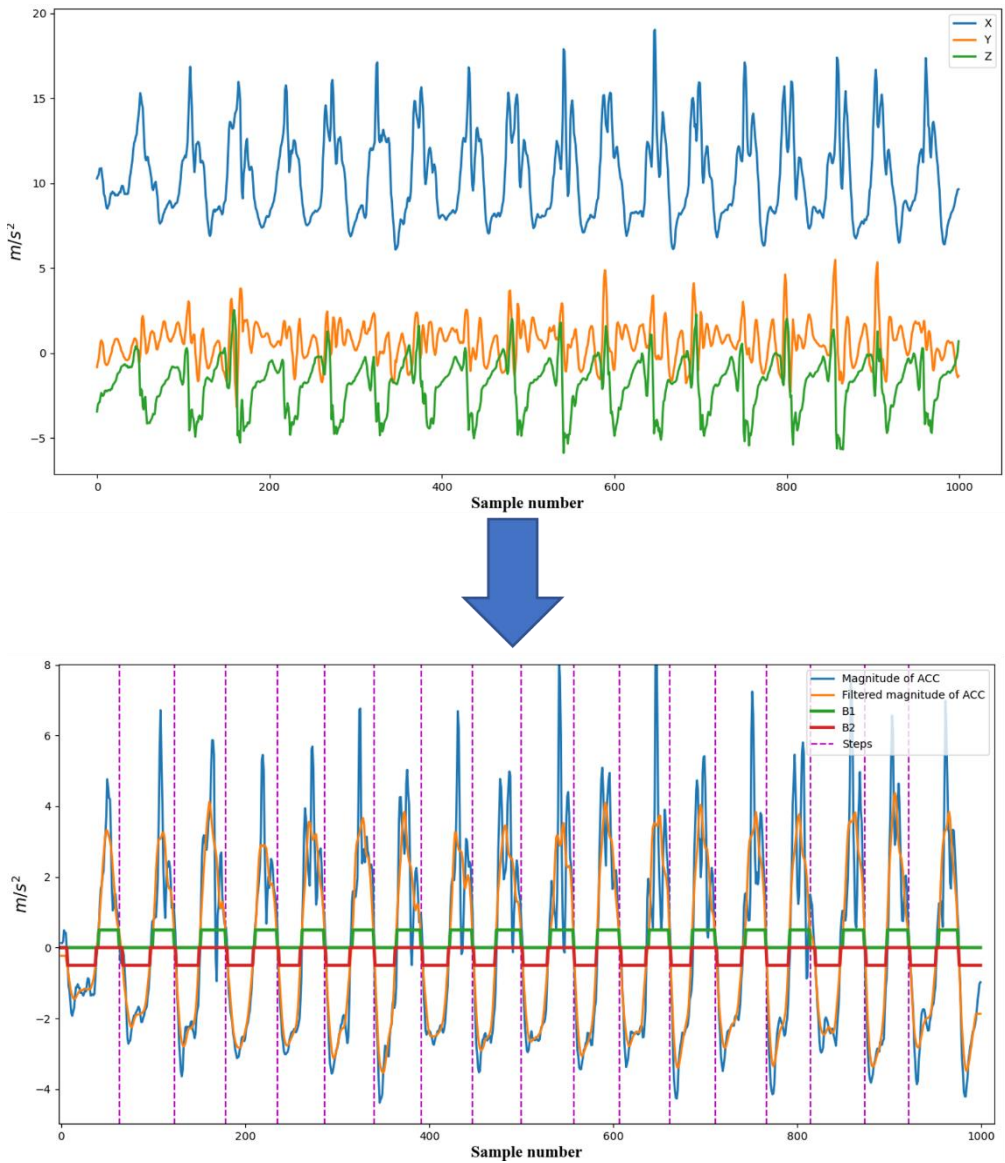

**Figure 1**. Step detection algorithm. ACC is the 3-axis acceleration signals, B1 and B2 are two series calculated by thresholds to detect walking swing and stance phase respectively, pink dash lines represent detected steps.

## Feature extraction

Since some gait metrics, such as stride length and body sway, are hard to be precisely extracted from acceleration signals, gait features extracted in this study were based on gait cadence and gait force. Gait cadence is the rate at which the individual feet contact the ground [34], which changes over time during daily-life walking. Therefore, we used not only the step count in one minute (the measurement of gait cadence in the laboratory gait tests [17]) but also the median of gait cycles and parameters in the frequency domain to describe gait cadence in this paper. Gait force reflects the ground reaction force during walking [35].

We extracted two categories of gait features: short-term (laboratory) gait features and daily-life gait features. Five short-term gait features were extracted from 1-minute laboratory walking tests of the LTMM dataset or 1-minute continuous walking segments (defined later) in both datasets to reflect the average gait cadence and gait force in that minute. Twenty daily-life gait features were extracted from 3 days after the enrollment for the LTMM dataset and 14 days before each PHQ-8 for the RADAR-MDD-KCL dataset to describe the distribution and variance of gait cadence and gait force over the period. Table 1 summarizes all gait features extracted in this paper.

**Short-term (laboratory) gait features**

For 1-minute acceleration signals, we first applied the step detection algorithm to obtain the gait cycles series ($Cycles$) and gait force series ($Force$). The median of gait cycles series and the number of steps were used to reflect the gait cadence of this minute from the time domain, which were denoted as *Median_Cycle* and *Step_Count,* respectively. To assess the gait cadence from the frequency domain, the power spectral density (PSD) of walking was obtained by applying the fast Fourier transformation (FFT) to the filtered magnitude ($\overline{r_i}$) of the acceleration signals. The peak frequency [43] and mean frequency [44] of 0.5-3 Hz band [29] of the PSD were used to reflect the main rhythm and average rhythm of steps from the frequency domain, which were denoted as *Peak_Freq* and *Mean_Freq,* respectively. For gait force, we calculated the median of $Force$ series (*Median_Force*) to represent the average power of all steps in one minute.

**Daily-life gait features**

Extracting daily-life gait characteristics from a relatively long-term period (3 days for LTMM or 14 days for RADAR-MDD-KCL) acceleration signals can be divided into 2 steps: (1) continuous walking segments detection and (2) daily-life gait feature extraction from detected continuous walking segments. A schematic diagram of daily-life gait feature extraction is shown in Figure 2. For the first step, the size of the walking segments was chosen to be 1 minute [45]. We applied the step detection algorithm to every minute of the long-term acceleration signals. The walking time (sum of all gait cycles in the minute) of a minute can reflect whether the participant is continuously walking in this minute. Intermittent walking (such as walking in a crowded environment or a walking-rest transition status) with a short walking time of a minute may not fully reflect a participant's normal walking patterns. Therefore, we set 50 seconds as the threshold for selecting continuous walking segments, that is, the segments with more than 50 seconds of walking time were selected for further analysis. In this step, depression score records with no continuous walking segments detected in the corresponding period were discarded.

In the second step, we first extracted 5 short-term (laboratory) gait features (described above) from each detected continuous walking segment. Then, for each short-term feature, we calculated 4 statistical second-order features (25th percentile, median, 75th percentile, and standard deviation) on all continuous walking segments detected over the long-term period (3 or 14 days) of each participant. For example, the 25th percentile of all continuous walking segments' *Median_Cycle* values over a long-term period was denoted as *Median_Cycle_25*, which was used to represent the gait cycle for faster steps (shorter gait cycle represent faster gait cadence) in a long-term period. A total of 20 statistical second-order features, called daily-life gait features, were extracted to reflect distribution (25th percentile, median, and 75th percentile) and variance of gait cadence and gait force over a long-term period corresponding to the depression score.

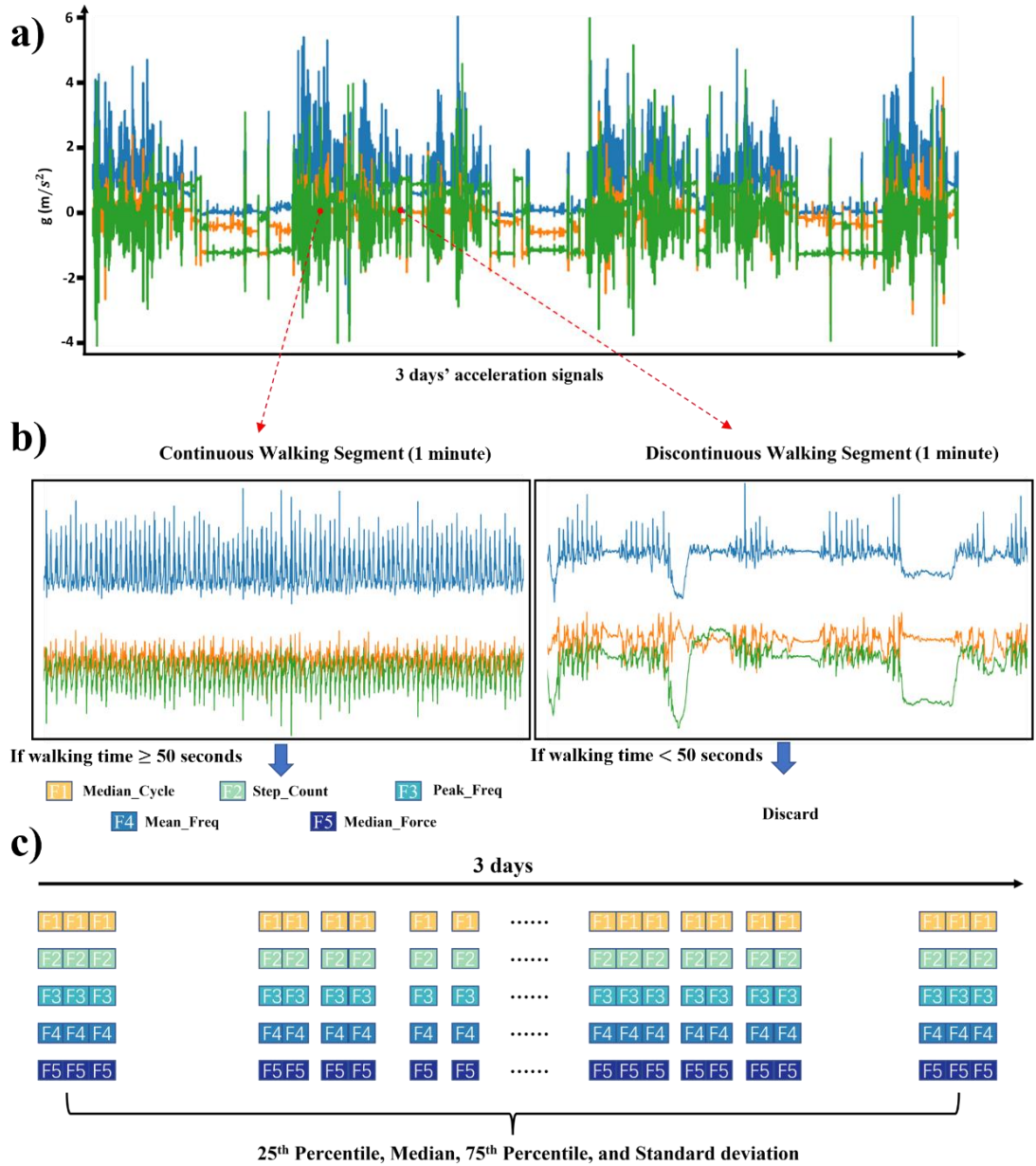

**Figure 2.** A schematic diagram of daily-life gait feature extraction for the Long-Term Movement Monitoring dataset. a) 3-axis acceleration signals of 3 consecutive days; b) examples of continuous and discontinuous walking segments and 5 short-term gait features (definitions in Table 1) were extracted from each continuous walking segment; c) daily-life gait feature extraction: 25th percentile, median, 75th percentile, and standard deviation of short-term gait feature values of all continuous walking segments over 3 days for each participant.

**Table 1.** A list of short-term (laboratory) gait features and daily-life gait features extracted in this paper and their short descriptions.

| Gait feature | Description |
|---|---|
| **Short-term (laboratory) gait feature** | |
| Median_Cycle | The median of gait cycles in the 1-minute walking. |
| Step_Count | The number of steps detected in the 1-minute walking. |
| Peak_Freq | The peak frequency in the PSD[a] of magnitude of 1-minute acceleration signals. |
| Mean_Freq | The mean frequency in the PSD of magnitude of 1-minute acceleration signals. |
| Median_Force | The median of gait force in the 1-minute walking. |
| **Daily-life gait feature** | |
| Median_Cycle_25 | The 25th percentile of median gait cycle values of all walking segments[b]. |
| Median_Cycle_50 | The median of median gait cycle values of all walking segments. |
| Median_Cycle_75 | The 75th percentile of median gait cycle values of all walking segments. |
| Median_Cycle_Std | The standard deviation of median gait cycle values of all walking segments. |
| Step_Count_25 | The 25th percentile of step count values of all walking segments. |
| Step_Count_50 | The median of step count values of all walking segments. |
| Step_Count_75 | The 75th percentile of step count values of all walking segments. |
| Step_Count_Std | The standard deviation of step count values of all walking segments. |
| Peak_Freq_25 | The 25th percentile of peak frequency values of all walking segments. |
| Peak_Freq_50 | The median of peak frequency values of all walking segments. |
| Peak_Freq_75 | The 75th percentile of peak frequency values of all walking segments. |
| Peak_Freq_Std | The standard deviation of peak frequency values of all walking segments. |
| Mean_Freq_25 | The 25th percentile of mean frequency values of all walking segments. |
| Mean_Freq_50 | The median of mean frequency values of all walking segments. |
| Mean_Freq_75 | The 75th percentile of mean frequency values of all walking segments. |
| Mean_Freq_Std | The standard deviation of mean frequency values of all walking segments. |
| Median_Force_25 | The 25th percentile of median gait force values of all walking segments. |
| Median_Force_50 | The median of median gait force values of all walking segments. |
| Median_Force_75 | The 75th percentile of median gait force values of all walking segments. |
| Median_Force_Std | The standard deviation of median gait force values of all walking segments. |

[a] PSD: power spectral density from 0.5 Hz to 3 Hz.
[b] All 1-minute continuous walking segments (defined in the Method section) in a specified time window (3 days for the Long-Term Movement Monitoring dataset and 14 days for the RADAR-MDD-KCL which is a subset of Remote Assessment of Disease and Relapse – Major Depressive Disorder dataset, which was collected from King's college London, United Kingdoms).

## Statistical Analyses

### Associations analyses

For the LTMM dataset, Spearman's coefficients [46] were performed to assess associations between the GDS-15 score and gait features (5 laboratory gait features and 20 daily-life features). As the data in the RADAR-MDD-KCL dataset is longitudinal (repeated PHQ-8 measurements for each participant), a series of pairwise linear mixed-effect regression models [47] with random participant intercepts were performed to explore the association between the PHQ-8 score and each of 20 daily-life gait features (no laboratory tests in the RADAR-MDD-KCL dataset). Age, gender, and the number of comorbidities were considered as covariates. The Benjamini-Hochberg method was used for multiple comparison corrections in both datasets [48].

**Multivariate linear (mixed-effect) regression models and likelihood ratio tests**

To test whether daily-life gait features can provide additional value for fitting depression scores (GDS-15 or PHQ-8), we built 2 nested regression models with and without daily-life gait features for both the LTMM and the RADAR-MDD-KCL datasets. Predictors of the model without daily-life gait features are age, gender, the number of comorbidities (only for the RADAR-MDD-KCL), and laboratory gait features (only for the LTMM). Since the data in the LTMM and RADAR-MDD-KCL datasets are cross-sectional and longitudinal, respectively, we used the multiple linear regression models in the LTMM dataset and multivariate linear mixed-effects regression models in the RADAR-MDD-KCL dataset. To avoid overfitting and multicollinearity issues, daily-life gait features were selected by a stepwise regression method (forward selection [49]) before including in the regression model with daily-life gait features. To indicate the proportion of data was explained by regression models, we calculated $R^2$ and adjusted $R^2$ for multiple linear regression models. For linear mixed-effect regression models, we calculated marginal $R^2$ for indicating data variance explained by fixed effects and conditional $R^2$ for representing data variance explained by both fixed effects and random effects [50]. The likelihood ratio tests [51] were performed to test whether the models with daily-life gait features fit the depression score significantly better than the models without daily-life gait features.

# Results

## Data summary

The 71 participants in the LTMM dataset have a mean (SD) age of 78.36 (4.71) years with 18 (25.35%) participants having potential depressive disorders (GDS-15≥ 5) with 69.82 (9.65) hours acceleration signals per participant. The RADAR-MDD-KCL dataset, after excluding PHQ-8 records without any continuous walking segment, contains 659 PHQ-8 records collected from 215 participants and corresponding 99445 hours (average 463 hours per participant). The cohort in the RADAR-MDD-KCL dataset has a mean (SD) age of 43.36 (15.12) years with the majority being females (75.35%) and a half of PHQ-8 records (50.08%) indicated potential depression symptoms (PHQ-8 ≥ 10). The amount of excluded data was recorded in Supplement Table 2. The average missing rate of acceleration signals collected by phones in the RADAR-MDD-KCL (70.60%) is significantly higher than acceleration signals collected by the wearable device in the LTMM dataset (3.03%). A summary of the demographics and distributions of depression scores and available acceleration signals for participants in the LTMM and the RADAR-MDD-KCL datasets is shown in Table 2.

**Table 2.** A summary of the demographics and depression score distribution of participants in the Long-term Movement Monitoring (LTMM) dataset and the RADAR-MDD-KCL[a] dataset.

| Characteristic | LTMM | RADAR-MDD-KCL |
|---|---|---|
| Participants, n | 71 | 215 |
| Age (years), mean (SD) | 78.36 (4.71) | 43.36 (15.12) |
| Female, n (%) | 46 (64.79%) | 162 (75.35%) |
| Depression score (GDS-15[b] and PHQ-8[c]), mean (SD) | GDS-15: 3.18 (2.81) | PHQ-8: 9.67 (5.84) |
| Potential depressive episode (GDS-15 $\geq$ 5 and PHQ-8 $\geq$ 10), n (%) | 18 (25.35%) | 330 (50.08%) |
| Number of completed depression questionnaires | 71 | 659[d] |
| Number of completed depression questionnaires for each participant, mean (SD) | 1 (0) | 3.09 (2.76) |
| Length of total available acceleration signals | 4817 hours | 99445 hours |
| Time window for daily-life gait feature[e] extraction | 3 days (72 hours) | 14 days (336 hours) |
| Length of available acceleration signals (hours) for each GDS-15/PHQ-8 record[f], mean (SD) | 69.82 (9.65) | 98.77 (105.20) |
| Average missing rate of acceleration signals (%) | 3.03% | 70.60% |
| Number of continuous walking segments[g] detected for each GDS-15/PHQ-8 record, mean (SD) | 73.48 (66.98) | 113.24 (170.48) |

[a] RADAR-MDD-KCL: A subset of the Remote Assessment of Disease and Relapse – Major Depressive Disorder dataset which was collected from King's College London, United Kingdom.
[b] GDS-15: 15-item Geriatric Depression Scale.
[c] PHQ-8: 8-item Patient Health Questionnaire.
[d] RADAR-MDD-KCL dataset has multiple PHQ-8 records for each participant which was conducted every two weeks.
[e] Daily-life gait feature was defined as the walking patterns extracted from acceleration

signals in real-world settings (not laboratory settings) in the Method section.

[f] We regarded acceleration signals in the two weeks before a PHQ-8 record as belonging to this PHQ-8 period. For the GDS-15 record, we considered acceleration signals of all 3-day activities belonging to this GDS-15 score.

[g] Continuous walking segment was defined as 1-minute acceleration signals with at least 50 seconds walking (See Method section).

## Associations between gait features and the GDS-15 in the LTMM dataset

The significant Spearman correlations between the GDS-15 score and gait features in the LTMM dataset are shown in Table 3. For laboratory gait features extracted from 1-minute laboratory walking tests, 4 of 5 gait features were significantly associated with the GDS-15 score. The results illustrated that the gait cadence and gait force in laboratory walking tests were significantly and negatively associated with the GDS-15 score. For daily-life gait features extracted from 3-day acceleration signals, 6 of 20 features were significantly associated with the GDS-15 score. The variance of gait cadence (*Step_Count_Std* and *Mean_Freq_Std*) and gait force (*Median_Force_Std*) over 3 days were significantly and negatively associated with the GDS-15 score. Notably, the GDS-15 score was significantly and negatively associated with gait cadence of faster steps over 3 days (75 percentile of gait frequency and 25 percentile of median gait cycles both represent faster steps).

**Table 3.** The significant Spearman correlations between the 15-item Geriatric Depression Scale score and gait features, including laboratory and daily-life gait features, in the Long-Term Movement Monitoring dataset.

| Feature[a] | $r_s$ | Adjusted P-value[b,c] |
|---|---|---|
| Laboratory gait features extracted from 1-minute laboratory walking acceleration signals | | |
|     Median_Cycle | 0.389 | .001 |
|     Step_Count | -0.372 | .002 |
|     Peak_Freq | -0.319 | .01 |
|     Median_Force | -0.250 | .04 |
| Daily-life gait feature extracted from 3-day activity acceleration signals | | |
|     Median_Cycle_25 | 0.309 | 0.01 |
|     Step_Count_Std | -0.305 | 0.01 |
|     Peak_Freq_75 | -0.265 | 0.03 |
|     Mean_Freq_75 | -0.248 | 0.04 |
|     Mean_Freq_Std | -0.26 | 0.03 |
|     Median_Force_Std | -0.296 | 0.02 |

[a] Definitions of gait features in this table are shown in Table 1.
[b] P values were adjusted by the Benjamini-Hochberg method for correction of multiple comparisons.
[c] Only significant correlations (adjusted P value <.05) are reported. Results for all gait features are shown in Supplement Table 3.

**Associations between daily-life gait features and PHQ-8 score in the RADAR-MDD-KCL dataset**

The pairwise linear mixed-effect models performed on the RADAR-MDD-KCL dataset revealed that 3 of 20 daily-life gait features extracted from 14-day acceleration signals were significantly associated with the PHQ-8 score (Table 4). Specifically, gait features (*Peak_Freq_75* and *Mean_Freq_75*) of the 75th percentile of gait frequency values over all continuous walking segments in 14 days were significantly and negatively associated with the PHQ-8 score. Similarly, the 25th percentile of gait cycles values over 14 day's continuous walking segments (*Median_Cycle_25*) was significantly and positively associated with the depression symptom severity. Notably, these 3 gait features were all related to faster steps over 14 days.

**Table 4.** Coefficient estimates, standard error (SE), t-test statistics, and adjusted P values from pairwise linear mixed-effect models[a] for exploring associations between daily-life gait features and depression symptom severity (8-item Patient Health Questionnaire) in the RADAR-MDD-KCL dataset[a].

| Feature[b] | Estimates | SE | t-value | Adjusted P-values[c,d] |
|---|---:|---:|---:|---:|
| Median_Cycle_25 | 0.094 | 0.048 | 1.943 | 0.04 |
| Peak_Freq_75 | -2.162 | 0.887 | -2.438 | 0.02 |
| Mean_Freq_75 | -3.215 | 1.228 | -2.617 | 0.01 |

[a] RADAR-MDD-KCL: A subset of Remote Assessment of Disease and Relapse – Major Depressive Disorder which was collected from King's college London.

[b] Definitions of daily-life gait features in this table are shown in Table 1.

[c] P values were adjusted by the Benjamini-Hochberg method for correction of multiple comparisons.

[d] Only significant correlations (adjusted P value <.05) are reported. Results for all gait features are shown in Supplement Table 4.

## Multiple linear regression models and likelihood-ratio test in the LTMM dataset

Table 5 represents the results of 2 nested multiple linear regression models with and without daily-life gait features in the LTMM dataset. There were 8 of 20 daily-life gait features were selected in the model with daily-life gait features using the forward selection method. The model with daily-life gait features achieved better performance ($R^2 = 0.42$, adjusted $R^2 = 0.22$) than the model without daily-life gait features ($R^2 = 0.09$, adjusted $R^2 = -0.03$). The likelihood ratio test showed that additional 8 daily-life gait features can significantly improve the regression model ($\chi^2 = 26.91 > \chi^2_{0.05}(8)$, P< .001).

**Table 5.** Results and performance of 2 nested multiple linear regression models with and without daily-life gait features in the Long-Term Movement Monitoring dataset.

| Feature[a] | Model with daily-life gait feature | | | Model without daily-life gait feature | | |
|---|---|---|---|---|---|---|
| | Estimate (SE[b]) | t value | P value | Estimate (SE) | t value | P value |
| (Intercept) | -9.30(19.83) | -0.47 | 0.64 | 20.54(15.60) | 15.60 | 0.19 |
| Age | -0.08(0.08) | -0.98 | 0.33 | -0.08(0.09) | 0.09 | 0.37 |
| Gender | -0.02(0.78) | -0.02 | 0.98 | -0.28(0.86) | 0.86 | 0.75 |
| Median_Cycle | 0.05(0.13) | 0.41 | 0.68 | -0.12(0.14) | 0.14 | 0.39 |
| Step_Count | -0.07(0.13) | -0.51 | 0.62 | -0.17(0.15) | 0.15 | 0.26 |
| Peak_Freq | -4.28 (2.34) | -1.83 | 0.07 | -1.41(2.54) | 2.54 | 0.58 |
| Mean_Freq | 10.18(3.99) | 2.55 | 0.01 | 4.23(3.79) | 3.79 | 0.27 |
| Median_Force | -0.81(2.08) | -0.39 | 0.70 | -2.41(2.24) | 2.24 | 0.29 |
| Median_Cycle_75 | 0.21(0.09) | 2.19 | 0.03 | —[c] | — | — |
| Median_Cycle_Std | -0.10(0.07) | -1.46 | 0.15 | — | — | — |
| Step_Count_50 | 0.16 (0.07) | 2.24 | 0.03 | — | — | — |
| Peak_Freq_25 | 16.30(7.82) | 2.09 | 0.04 | — | — | — |
| Peak_Freq_50 | -9.12(6.32) | -1.44 | 0.16 | — | — | — |
| Peak_Freq_75 | -6.34(4.43) | -1.43 | 0.16 | — | — | — |
| Mean_Freq_75 | -11.60(5.58) | -2.08 | 0.04 | — | — | — |
| Mean_Freq_Std | 28.30(13.62) | 2.08 | 0.04 | — | — | — |
| $R^2$ | 0.42 | | | 0.09 | | |
| Adjusted $R^2$ | 0.22 | | | -0.03 | | |
| LR test[e]: $\chi^2$ | 26.91 | | | | | |
| LR test: P value | <0.001 | | | | | |

[a] Definitions of gait features in this table are shown in Table 1. The daily-life gait features used in the model with daily-life gait features were selected by the forward selection method. Median_Cycle, Step_Count, Peak_Freq, Mean_Freq, Median_Force are 5 laboratory gait features extracted from 1-minute laboratory walking tests.
[b] SE: standard error.
[c] Not applicable.
[e] The critical value of the likelihood ratio statistic: $\chi^2_{0.05}(8) = 15.51$.

## Multivariate linear mixed-effect regression models and likelihood-ratio test in the RADAR-MDD-KCL dataset

The results of 2 nested multivariate linear mixed-effect regression models with and without daily-life gait features of the RADAR-MDD-KCL dataset are displayed in Table 6. There were 4 daily-life gait features were selected in the model with daily-life gait features by the stepwise regression method. The model with daily-life gait features achieved better performance (Marginal $R^2$=0.22 and Conditional $R^2$= 0.76) than the model without daily-life gait features (Marginal $R^2$=0.16 and Conditional $R^2$= 0.72). The likelihood ratio test showed the included daily-life gait features can significantly improve the regression model ($\chi^2 = 16.16 > \chi^2_{0.05}(4)$, P= .002).

**Table 6.** Results and performance of multivariate linear mixed-effect regression models with and without daily-life gait features in the RADAR-MDD-KCL dataset.

| Feature[a] | Model with daily-life gait features | | | Model without daily-life gait features | | |
| --- | --- | --- | --- | --- | --- | --- |
| | Estimate (SE[b]) | t value | P | Estimate (SE) | t value | P |
| (Intercept) | 16.32 (2.36) | 6.93 | <0.001 | 13.00 (1.28) | 10.19 | <0.001 |
| Age | -0.12 (0.02) | -5.11 | <0.001 | -0.11(0.02) | -4.87 | <0.001 |
| Gender | -0.26 (0.87) | -0.30 | 0.77 | -0.40 (0.82) | -0.49 | 0.63 |
| Comorbidity[d] | 1.11 (0.24) | 4.60 | <0.001 | 1.30 (0.22) | 6.01 | <0.001 |
| Step_Count_50 | -0.13 (0.04) | -2.91 | 0.004 | —[c] | — | — |
| Step_Count_75 | 0.12 (0.04) | 2.81 | 0.005 | — | — | — |
| Mean_Freq_25 | 6.55 (1.92) | 3.41 | <0.001 | — | — | — |
| Mean_Freq_75 | -7.77 (2.10) | -3.69 | <0.001 | — | — | — |
| Marginal $R^2$ | 0.22 | | | 0.16 | | |
| Conditional $R^2$ | 0.76 | | | 0.72 | | |
| LR test[e]: $\chi^2$ | 16.16 | | | | | |
| LR test: P value | 0.002 | | | | | |

[a] Definitions of gait features in this table are shown in Table 1. The daily-life gait features used in the model with daily-life gait features were selected by the forward selection method.
[b] SE: standard error.
[c] Not applicable.
[d] The number of comorbidities that listed in the Supplement Table 1.
[e] The critical value of the likelihood ratio statistic: $\chi^2_{0.05}(4) = 9.49$.

## Discussion

### Principal findings

This study explored the associations between depression symptom severity and daily-life gait characteristics of real-world settings using two separate datasets (LTMM and RADAR-MDD-KCL) from different populations, assessed by different depression questionnaires and accelerometer devices. To the best of our knowledge, our study is the first to investigate associations between depression symptom severity and daily-life gait characteristics derived from acceleration signals in real-world scenarios. We extracted 20 daily-life gait features to describe the distribution (25th percentile, median, and 75th percentile) and variance of gait cadence and force over a long-term period corresponding to a self-reported depression score. The main findings of this paper are 1) gait cadence of faster steps (75th percentile) over a long-term period has a significant negative association with the depression symptom severity of that period, 2) daily-life gait features could provide additional value for evaluating depression symptom severity relative to laboratory gait characteristics and demographics, and 3) wearable devices and mobile phones both have potentials to capture the associations between daily gait and depression.

**Associations between depression symptom severity and gait features**

The results of Spearman correlations between laboratory gait features and the GDS-15 score in the LTMM dataset are consistent with previous studies [17-25], that is, the participants with more severe depression symptoms were likely to have slower gait cadence (longer the median of gait cycles and lower gait frequency) and smaller gait force in laboratory walking tests. For daily-life gait features, gait cadence of faster steps (75th percentile) over a long-term period has a significant negative association with the depression symptom severity of that period. Specifically, if a participant has severe depression symptoms, the frequency of his/her faster steps (over a specific period) is lower (or the median of these steps' cycles is longer). This finding is consistent in both the LTMM and the RADAR-MDD-KCL datasets. As the situations in daily-life walking are complex (such as walking during the day or at night, walking under fatigue or walking after rest, and walking to a destination or navigating a crowded supermarket), the performances of participants during walking were also different [28]. Therefore, from the main finding of this paper, we speculated that the faster steps over a long-term period could represent the optimal performance of a participant which could be associated closely with their depression status. A previous study of another field (fall risks) also showed that the extreme values (10th and 90th percentile) of gait characteristics could reflect the physical or mental conditions better than the median value of gait characteristics [28].

For gait features related to the variance of gait over a long-term period, associations with the depression score were inconsistent across two datasets. In the LTMM dataset, we found features of the variance of gait in 3 days were significantly and negatively associated with the depression symptom severity (Table 3) indicating that participants with lower depression symptom severity (GDS-15 score) tend to have higher variance in daily-life gaits' cadence and force. In contrast, participants with severe depression symptoms were likely to have relatively monotonous walking over 3 days. However, these features were not significantly associated with the PHQ-8 score in the RADAR-MDD-KCL dataset. One reason is that the magnitude of the acceleration signals depends on the location of the accelerometers attached to the body [52]. As acceleration signals in the RADAR-MDD-KCL were collected by mobile phones, the variable locations of phones when attached to participants' bodies (such as in the hand, handbag, and pocket) affected the amplitudes of acceleration signals. Therefore, the magnitude of acceleration signals cannot fully reflect the gait force and the changes in the magnitude of acceleration signals. Another potential reason is that the missing rate of acceleration data is relatively high in the RADAR-MDD-KCL dataset, by comparison to the LTMM dataset (Table 2), so the features of standard deviation cannot fully reflect the variance of participants' steps over 14 days.

## Additional value of daily-life gait characteristics for evaluating the depression symptom severity

The results of likelihood ratio tests for regression models with and without daily-life gait features in the LTMM and the RADAR-MDD-KCL datasets (Table 5 and Table 6) both indicated that daily-life gait characteristics could provide additional value for evaluating depression symptom severity relative to laboratory gait features (only in LTMM) and demographics in two datasets.

From the results of multiple linear regression models in the LTMM dataset (Table 5), we noticed that laboratory gait features and demographics only explained a small proportion of data variance ($R^2 = 0.09$), while included daily-life gait features explained an extra 33% data variance ($R^2 = 0.42$). This finding indicated that the laboratory walking test may be affected by several factors, such as subjective psychological factors and laboratory-controlled conditions, which may not fully reflect the condition of a participant's mental health [27, 28]. From the results of multivariate linear mixed-effect regression models in the RADAR-MDD-KCL dataset (Table 6), we found that although additional daily-life gait features explained an extra 6% data variance relative to demographics, random effects (random intercepts for participants) explained more data variance than fixed effects (daily-life gait features, age, gender, the number of comorbidities) (conditional $R^2$=0.76 and marginal $R^2 = 0.22$). This indicated that variance in the within-individual level is relatively small compared with the between-participant level in the RADAR-MDD-KCL dataset. One potential reason is that the average number of PHQ-8 records of each participant included in gait analysis is relatively small (3.09 PHQ-8 records).

## The LTMM and RADAR-MDD-KCL datasets

Populations, depression questionnaires, and accelerometer devices are all different across the LTMM and the RADAR-MDD-KCL datasets. Since the LTMM dataset was designed for exploring elderly people's fall risks, the participant's depression status was not considered in the study protocol. The proportion of participants with potential depressive disorders was less than one-third (25.35%), which may lead to statistical bias. To the best of our knowledge, existing studies on the LTMM dataset were all related to fall risks, therefore, this paper is the first to use the LTMM dataset to explore the associations between depression and gait. Compared with the LTMM dataset, the RADAR-MDD-KCL dataset has a wider age distribution population with longer follow-up. Participants in the RADAR-MDD-KCL dataset have at least one diagnosis of depression in the last 2 years, which may be the reason that the RADAR-MDD-KCL dataset has more than half of PHQ-8 records with potential severe depression symptoms (50.08%; Table 2). Furthermore, as the RADAR-MDD-KCL dataset has multilevel data, the results indicated the significant associations between daily-life gait patterns and depression exist in both individual and population levels. Some consistent results in two separate ambulatory datasets indicated the generalizability of our findings, which provide additional confidence in associations between depression and daily-life gait and features identified in this paper.

**Limitations**

In this work, we did not fully explore all the gait characteristics of daily-life walking, because some gait metrics are hard to be extracted from acceleration signals such as step length, body posture, and body swing [14]. Gait features used in this paper were only related to gait cadence and gait force. Our aim was to explore information from daily-life walking that can improve the evaluation of depression symptom severity, rather than replacing the "golden standard" laboratory gait assessment. Therefore, we focused on illustrative statistics-based features demonstrating the importance of long-term gait features for depression monitoring. More complex features such as nonlinear features will be considered in future research.

In our previous studies on sleep and Bluetooth data in the RADAR-MDD dataset [53, 54], each participant has an average of 8 PHQ-8 records with sleep or Bluetooth data. However, each participant has only an average of 3.09 PHQ-8 records with detected continuous walking segments (Table 2). Possible reasons are that high battery consumption and network traffic for uploading the raw acceleration signals, the Android operating system moderation of resources, and the range of Android phones have some variability in the performance of accelerometer sensors. We also found many PHQ-8 records with acceleration signals have no continuous walking segment detected (Supplement Table 2). The potential reasons are that several participants have low mobility because of physical morbidities and depression [55], and some participants may not bring the phone during their walking.

Although missing acceleration signals caused several steps to be undetected and affected the distribution and variance of steps, the remaining detected continuous walking segments can still partially reflect the gait cadence and force of the participant.

Since we are not measuring mobility and activity but gait characteristics, we included PHQ-8 records with at least one continuous walking segment into data analyses. The results also showed that the associations between daily-life walking patterns and the depression score could still be captured by the mobile phone's acceleration signals.

According to the findings in this paper, we can consider uploading gait cycles instead of uploading raw acceleration signals in future long-term monitoring research, which may spend less battery consumption and reduce the missingness of steps information. It is not difficult to implement, as most current smartphones have real-time step detection functions or apps [56, 57].

Gait characteristics could be affected by some physical diseases, neurological disorders, and age [58-60]. Although we considered the number of comorbidities (Supplement Table1) and demographics as covariates to control the confounder factors, physical comorbidities and other comorbidities may have different impacts on the gait characteristics. We will consider a wider range of comorbidities and investigate them further in future research.

We considered a 1-minute acceleration segment with more than 50 seconds of walking as a continuous walking segment and included it in our analysis. The 1-minute segment size was suggested by previous studies [29, 45], and the threshold of 50 seconds for continuous walking was manually specified by our experience for excluding some abnormal steps in several conditions, such as walking in a crowded environment and intermittent walking indoors, which may not reflect the normal walking patterns. Another past study used a 10-second time window for the step detection and short-term feature extraction [28], but this may reduce the accuracy of step detection and involve abnormal steps. Therefore, we will explore the optimal window size for detecting steps and optimal threshold for continuous walking in future research.

## Conclusion

We found significant links between depression symptom severity and daily-life gait characteristics and these associations could be captured by acceleration signals of both wearable devices and mobile phones. These daily-life walking patterns could provide additional value for understanding depression manifestations relative to gait patterns in laboratory walking tests and demographics. The gait cadence of faster steps of daily-life walking over a long-term period has a significant and negative association with depression symptom severity of that period which has the potential to be a biomarker for detecting depression. This study illustrated that long-term gait monitoring for depression may contribute to clinical tools to remotely monitor mental health in real-world settings.

## Acknowledgments

Funding The RADAR-CNS project has received funding from the Innovative Medicines Initiative 2 Joint Undertaking under grant agreement No 115902. This Joint Undertaking receives support from the European Union's Horizon 2020 research and innovation programme and EFPIA (www.imi.europa.eu). This communication reflects the views of the RADAR-CNS consortium and neither IMI nor the European Union and EFPIA are liable for any use that may be made of the information contained herein. The funding body have not been involved in the design of the study, the collection or analysis of data, or the interpretation of data.

This study represents independent research part funded by the National Institute for Health Research (NIHR) Maudsley Biomedical Research Centre at South London and Maudsley NHS Foundation Trust and King's College London. The views expressed are those of the author(s) and not necessarily those of the NHS, the NIHR or the Department of Health and Social Care.

We thank all the members of the RADAR-CNS patient advisory board for their contribution to the device selection procedures, and their invaluable advice throughout the study protocol design.

This research was reviewed by a team with experience of mental health problems and their careers who have been specially trained to advise on research proposals and documentation through the Feasibility and Acceptability Support Team for Researchers (FAST-R): a free, confidential service in England provided by the National Institute for Health Research Maudsley Biomedical Research Centre via King's College London and South London and Maudsley NHS Foundation Trust.

We thank all GLAD Study volunteers for their participation, and gratefully acknowledge the NIHR BioResource, NIHR BioResource centres, NHS Trusts and staff for their contribution. We also acknowledge NIHR BRC, King's College London, South London and Maudsley NHS Trust and King's Health Partners. We thank the National Institute for Health Research, NHS Blood and Transplant, and Health Data Research UK as part of the Digital Innovation Hub Programme.

RADAR-MDD will be conducted per the Declaration of Helsinki and Good Clinical Practice, adhering to principles outlined in the NHS Research Governance Framework for Health and Social Care (2nd edition). Ethical approval has been obtained in London from the Camberwell St Giles Research Ethics Committee (REC reference: 17/LO/1154), in London from the CEIC Fundació Sant Joan de Deu (CI: PIC-128-17) and in The Netherlands from the Medische Ethische Toetsingscommissie VUms (METc VUmc registratienummer: 2018.012 – NL63557.029.17).

**Conflicts of Interest**

SV and VAN are employees of Janssen Research and Development LLC. PA is employed by the pharmaceutical company H. Lundbeck A/S. DCM has accepted honoraria and consulting fees from Apple, Inc, Otsuka Pharmaceuticals, Pear Therapeutics, and the One Mind Foundation; has received royalties from Oxford Press; and has an ownership interest in Adaptive Health, Inc. MH is the principal investigator of RADAR-CNS, a private public pre-competitive consortium which receives funding from Janssen, UCB, Lundbeck, MSD and Biogen.